\documentclass[a4paper,twoside]{article}

\usepackage{amsmath,amsfonts,amssymb}
\usepackage{cite}
\usepackage{graphicx}
\usepackage{color}
\usepackage{comment}
\usepackage{amsmath,amssymb}
\usepackage[utf8]{inputenc}
\usepackage{multirow}
\usepackage{booktabs}
\usepackage{floatrow}
\usepackage{algorithm}
\usepackage{algorithmic}
\usepackage{array}
\usepackage{fixltx2e}
\usepackage{dblfloatfix}
\usepackage{url}
\usepackage{caption}
\usepackage{subcaption}
\usepackage[colorlinks=true, allcolors=blue]{hyperref}
\usepackage{epsfig}
\usepackage{subcaption}
\usepackage{calc}
\usepackage{amssymb}
\usepackage{amstext}
\usepackage{amsthm}
\usepackage{multicol}
\usepackage{pslatex}
\usepackage{apalike}
\usepackage{SCITEPRESS}     

\begin{document}

\title{Roughness Index and Roughness Distance \\ for Benchmarking Medical Segmentation}

\author{\authorname{Vidhiwar Singh Rathour, Kashu Yamakazi and T. Hoang Ngan Le }
\affiliation{Department of Computer Science and Computer Engineering, University of Arkansas, Fayetteville, Arkansas USA 72701}
\email{\{vsrathou, kyamazak, thile\}@uark.edu}
}



\abstract{Medical image segmentation is one of the most challenging tasks in medical image analysis and has been widely developed for many clinical applications. Most of the existing metrics have been first designed for natural images and then extended to medical images. While object surface plays an important role in medical segmentation and quantitative analysis i.e. analyze brain tumor surface, measure gray matter volume, most of the existing metrics are limited when it comes to analyzing the object surface, especially to tell about surface smoothness or roughness of a given volumetric object or to analyze the topological errors.
In this paper, we first \textbf{analysis both pros and cons of all existing medical image segmentation metrics}, specially on volumetric data. We then propose an appropriate \textbf{roughness index and roughness distance} for medical image segmentation analysis and evaluation. Our proposed method addresses two kinds of segmentation errors, i.e. (i) \textbf{topological errors on boundary/surface}  and (ii) \textbf{irregularities on the boundary/surface}. The contribution of this work is four-fold: (i) detect irregular spikes/holes on a surface, (ii) propose roughness index to measure surface roughness of a given object, (iii) propose a roughness distance to measure the distance of two boundaries/surfaces by utilizing the proposed roughness index and (iv) suggest an algorithm which helps to remove the irregular spikes/holes to smooth the surface. Our proposed roughness index and roughness distance are built upon the solid surface roughness parameter which has been successfully developed in the civil engineering.}

\onecolumn \maketitle \normalsize \setcounter{footnote}{0} \vfill

\section{\uppercase{Introduction}}
\label{sec:introduction}
In this paper we first discuss the pros and cons of various metrics that have been commonly used for bench-marking the medical image segmentation task. We emphasize on the limitations of existing metrics, such as Hausdorff distance when evaluating the volumetric segmentation. Our study shows that the existing volumetric metrics are unable to measure the topological errors specially when irregular spikes/holes are on the surface. We then propose (i) an algorithm that helps to detect irregular spikes/holes that exist on a given object surface; (ii) a roughness index that describes how rough an object is given an object's surface; (iii) a roughness distance that aims at comparing the surfaces between two given objects; (iv) an algorithm that aims at removing the small outliers and the irregular spikes/holes to smooth the surface. As compared to other volumetric segmentation metrics i.e. Hausdorff distance, our proposed roughness distance is able to measure the topological error whereas roughness index evaluates the surface roughness. Furthermore, we conduct the experiment to show that our proposed irregular spikes/holes detection and surface smoothing can be applied as a post-processing step in any image segmentation algorithm to improve the accuracy.
\section{\uppercase{Description of purpose}}
\label{sec:descofpurpose}

\begin{figure}
\includegraphics[width=1\linewidth]{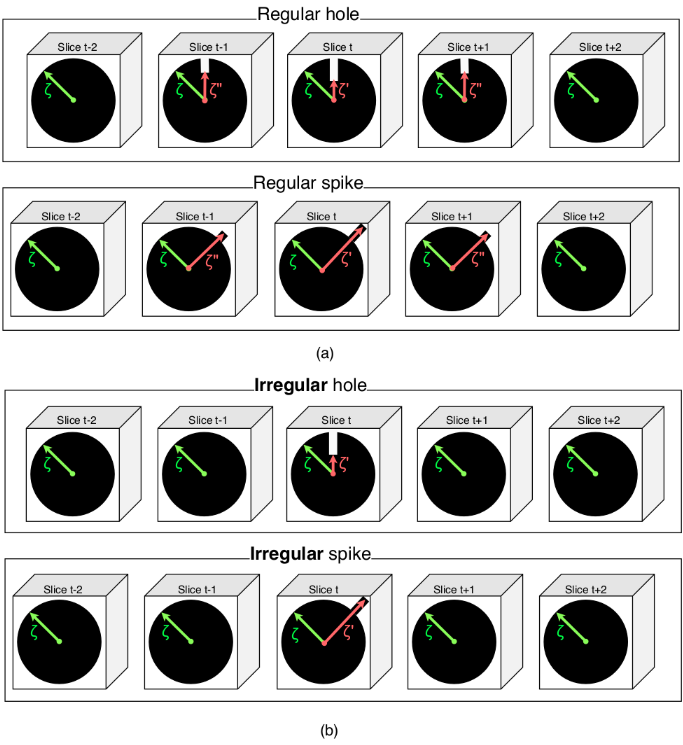}
	\caption{An illustration of a regular spike/hole (a) v.s an irregular spike/hole (b)}

	\label{Fig:motivation}
\end{figure}

Medical image segmentation is an important research topic in medical analysis and has attracted attention in past couple of years. With the abundance of medical data available it has become easier to perform segmentation task. However, evaluation and validation of medical segmentation, specially volumetric data is still a major concern because majority evaluation metrics have been developed as piece-wise setting for 2D natural images and then extended to medical images including volumetric data. As categorized in \cite{shi2013objective}, there are four types of segmentation errors i.e. \textit{quantitative or the number of objects, area of segmentation, contour or the object boundary, and the presence of holes, or irregularities in the boundary of segmentation}. The first type of error, which regards the number of objects, can be mitigated by increasing the training data. Most of the common evaluation metrics (i.e. Dice score , Sensitivity, Specificity, etc) have focused to solve the second type of error, i.e. area of segmentation which is a well-known problem in any segmentation task in both computer vision and medical analysis. For the third type of error, i.e. object contour/boundary error, there are a limited number of metrics that have been developed. Hausdorff distance (HDD) and Average  Symmetric  Surface  Difference (ASSD) \cite{gerig2001valmet} are the ones that have been used for calculating errors on object surface. The last error, which is related to topological errors such as holes and spikes, still remains as a challenging problem in medical analysis. Several attempts such as \cite{Surface_Constrained} \cite{surface_segmentation} \cite{smoothness_graph} has focused on the last error category by considering the smoothness and roughness criteria. \textbf{In this work, we address the last two kinds of errors, i.e. (i) topological error on boundary/surface  and (ii) irregularities on boundary/surface} as demonstrated in Fig:\ref{Fig:motivation}.

Different from 2D objects, volumetric objects need the \textit{consistency} and \textit{continuous} between slides. A comparison between consistency-inconsistency and regularity-irregularity in volumetric data is given in Fig:\ref{Fig:motivation} where each slide is presented in a cuboid (one volumetric is considered as a set of slices) and $\zeta$ is the distance between the surface and center of gravity. The \textit{inconsistency} or \textit{irregularity} is defined as an abrupt or a sudden spike/hole. In Fig:\ref{Fig:motivation}, the regular spike/hole is given in the top (Fig:\ref{Fig:motivation}.a) where spike or hole is gradually formed from slice to slice whereas the irregular spike/hole is given in the bottom (Fig:\ref{Fig:motivation}.b) where spike or hole suddenly appeared.



Different from the previous works \cite{Surface_Constrained} \cite{surface_segmentation} \cite{smoothness_graph} which use geometric graph i.e., minimum s-t cut, we make use of \textbf{solid surface roughness parameter in civil engineering to propose roughness metric} \cite{chang2006application} \cite{tonietto2019new} \cite{gadelmawla2002roughness}. Our contribution can be summarized as follows:
\begin{itemize}
    \item Revise and analyze the existing segmentation metrics that have been used in medical analysis (Sec:\ref{sec:relatedwork}).
    \item Propose an algorithm which helps to detect all irregular spikes/holes on the object surface (Sec:\ref{sec:detect}).
    \item Introduce a \textbf{roughness index} that measures the surface roughness given an object in (Sec:\ref{sec:ri}). Our proposed roughness index is based on the solid surface roughness parameter that has been successfully developed in the civil engineering \cite{chang2006application} \cite{tonietto2019new} \cite{gadelmawla2002roughness}.
    \item Propose a \textbf{roughness distance} metrics which computes the surface distance between two surfaces (Sec:\ref{sec:surfdist}). 
    \item Propose an algorithm which helps to remove the irregular spikes/holes and to smooth the contour (Sec:\ref{sec:smooth}).
\end{itemize}

\section{RELATED WORK}
\label{sec:relatedwork}
In this section, we will revise all existing segmentation metrics that have been commonly used in medical analysis. We first categorize the existing segmentation metrics into two groups, namely, region-based metrics and boundary-based metrics. We then analyse the pros and cons of each metric in the following subsections.


\subsection{Region-based metrics}
By definition, region-based metrics are used to evaluate the area occupied by the segmentation. The region-based metrics, which are based on pixel-wise, have been first developed for spatial images (2D) segmentation in computer vision in general and then extended to volumetric (3D) segmentation in medical imaging. These types of metrics tend to work well when there is clear demarcation with respect to data and when the contour is smooth. However they tend to fail when the the data has holes or boundary is irregular. These metrics tend to evaluate the second type of segmentation error, i.e. area of segmentation. The following is some common region based metrics that are popularly used volumetric segmentation. 

\noindent
\textbf{Segmentation Problem Setting}:
In the image segmentation problem, evaluation process is performed between the ground-truth G created by the human and segmentation predicted P by some algorithmic model.  

\noindent
\textbf{Dice Similarity Coefficient (DSC)}
Initially introduced as Dice \cite{dice1945measures} also known as the F1 score is one of the most commonly used metrics in validating medical image segmentation \cite{linguraru2012statistical} \cite{linguraru2009renal} in both spatial images and volumetric data. Lets consider $P$ as the predicted volumetric segmentation vector and $G$ as the ground-truth, then DSC can be calculated as shown in Eq:\ref{eq:DSC}.

\begin{equation}
\operatorname{DSC} = \frac{2|P \cap G|}{|P| + |G|}
\label{eq:DSC}
\end{equation}

\noindent
\textbf{Symmetric Volume Difference (SVD)} introduced by \cite{campadelli2009liver} and \textbf{Jaccard Similarity Coefficient (JSC)} introduced by \cite{liu2012effective} are similar to DSC and can be mathematically computed from DSC as shown in Eq:\ref{eq:SVD} and Eq:\ref{eq:JSC}.

\begin{equation}
\operatorname{SVD} = {1 - DSC}
\label{eq:SVD}
\end{equation}

\begin{equation}
\operatorname{JSC} = \frac{|P \cap G|}{|P \cup G|}= \frac{DSC}{2 - DSC}
\label{eq:JSC}
\end{equation}

DSC although works well with data that is clearly demarcated, yet it tends to produce unwanted results if the segmentation boundary is ambiguous. Also DSC cannot tell anything about the boundary information, roughness and smoothness of a volumetric surface or the topological error on the boundary surface. JSC and SVD have the same inherent problems as DSC. 

\noindent
\textbf{Precision (Pre), Recall (Rec) and Sensitivity (Sens)}
Precision is defined as the volume of correctly segmented volume to the total volume that has been segmented. Recall (also referred to as Sensitivity) is the the ratio of correctly segmented volume over the ground-truth.
\begin{equation}
\operatorname{Pre} = \frac{|P \cap G|}{|P|}
\label{eq:PRE}
\end{equation}
\begin{equation}
\operatorname{Rec/Sens} = \frac{|P \cap G|}{|G|}
\label{eq:REC}
\end{equation}
Precision takes into account only the volume that has been segmented correctly but does not consider the under-segmented volume. Recall on the other hand does not consider the over-segmented volume. However these two metrics are extensively being used in computer vision for segregation tasks \cite{wolz2012multi} \cite{campadelli2010segmentation}.

\noindent

\begin{figure}[!h]
	\includegraphics[width= 1\linewidth]{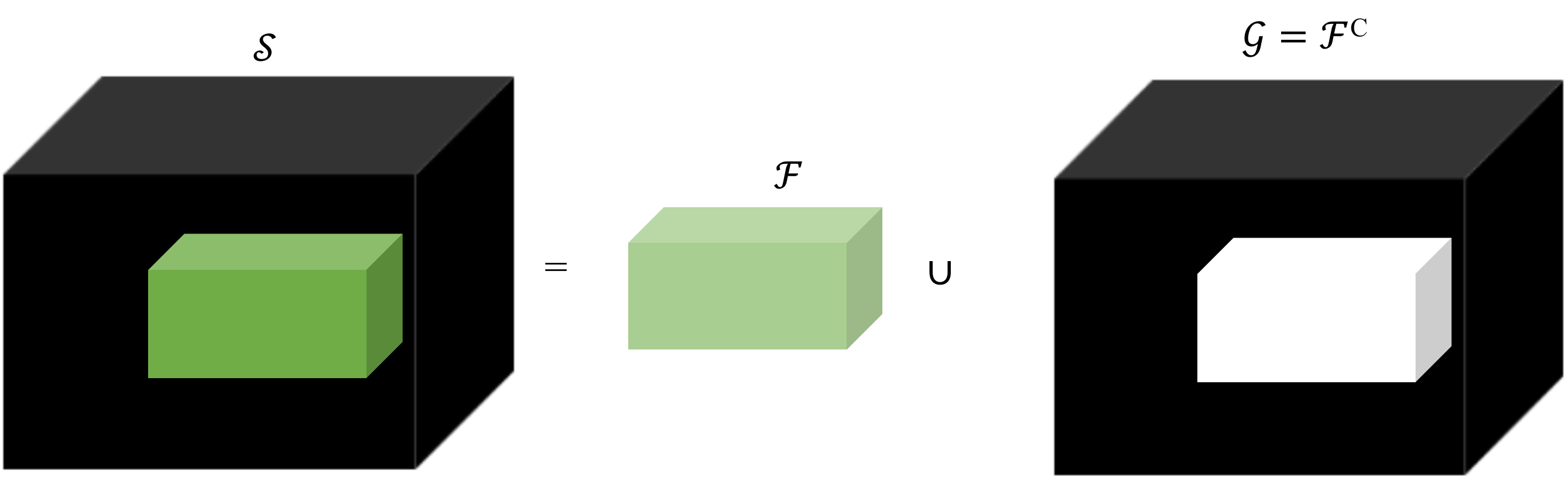}
	\caption{An Illustration of compliment using in Eq:\ref{eq:SET}. Green cuboid $F$ represents the set for which compliment is being calculated, and Black cuboid $S$ represents the universal set of which $F$ is a part.     } 
	\label{Fig:fig1}
\end{figure}

\textbf{Specificity (Spec)}
Specificity also referred to as Selectivity is the ratio of portion of total volume that is not common to the ground-truth ($G$) and predicted segmentation ($P$) by the portion not included in ground-truth ($G$).
True Negative (TN) is the portion of volume that is not common to the ground-truth ($G$) and predicted segmentation ($P$) and False Positive (FP) is the potion of volume belonging to predicted segmentation ($P$) that is not common to ground-truth ($G$):
\begin{equation}
\operatorname{Specificity(Spec)} = \frac{|(P \cup G)^C|}{|G^C|}
\label{eq:SET}
\end{equation}
Here $C$ denotes the compliment component which is illustrated in Fig:\ref{Fig:fig1}. The segmented volumetric $\mathcal{S}$ contains two parts corresponding to foreground $\mathcal{F}$ and background $\mathcal{G}$, where  $\mathcal{G} =  \mathcal{F}^C$

\noindent
\textbf{Relative Volume Difference (RVD)}
RVD is defined as the ratio of absolute difference in volume between the predicted volumetric segmentation vector ($P$) and the ground-truth ($G$)  to the ground-truth ($G$).
It is commonly used as a reference to other metrics\cite{heimann2009comparison} \cite{linguraru2012statistical}.
\begin{equation}
\operatorname{RVD} = |\frac{|P| - |G|}{|G|}|
\label{eq:RVD}
\end{equation}
RVD computed the relative difference in volume between predicted volumetric segmentation vector ($P$) and the ground-truth ($G$) and hence it does not take into consideration the overlap between them.

\subsection{Boundary-based metrics}

Different from region-based metrics, which are designed to work on entire area, boundary-based metrics focus on boundary or surface only. In this section, we revise two common boundary-based metrics, namely, Average Symmetric Surface Difference (ASSD) and Hausdorff Distance (HDD) as follows:

\noindent
\textbf{Hausdorff Distance (HDD)}

Hausdorff distance (HDD) is defined as the maximum possible distance from a point/voxel on one boundary/surface to the corresponding closest point/voxel on another boundary/surface \cite{gerig2001valmet} \cite{chen2012domain} \cite{liu2012effective} \cite{chen2012medical}. The HDD between the ground-truth boundary/surface $\partial G$ and the predicted segmentation boundary/surface $\partial P$ is defined as follows:
\begin{equation}
\operatorname{HDD} = \max_{x \in \partial G}((|x,\partial P|_{L2}))
\label{eq:HDD3}
\end{equation}
where $|x,\partial P|_{L2}$ is the shortest $L_2$ distance between a point/voxel $x$ on the ground-truth boundary/surface $\partial G$ and the predicted segmentation boundary/surface $\partial P$, namely, $|x,\partial P|_{L2} = \min_{y \in \partial P}||(x - y)||^2$. Thus, Eq.\ref{eq:HDD} is rewritten as:
\begin{equation}
\operatorname{HDD} = \max_{x \in \partial G}((|x,\partial P|_{L2})) = \max_{x \in \partial G}((\min_{y \in \partial P}||x - y||^2))
\label{eq:HDD}
\end{equation}

\begin{figure}[!h]
	\includegraphics[width= 1\linewidth]{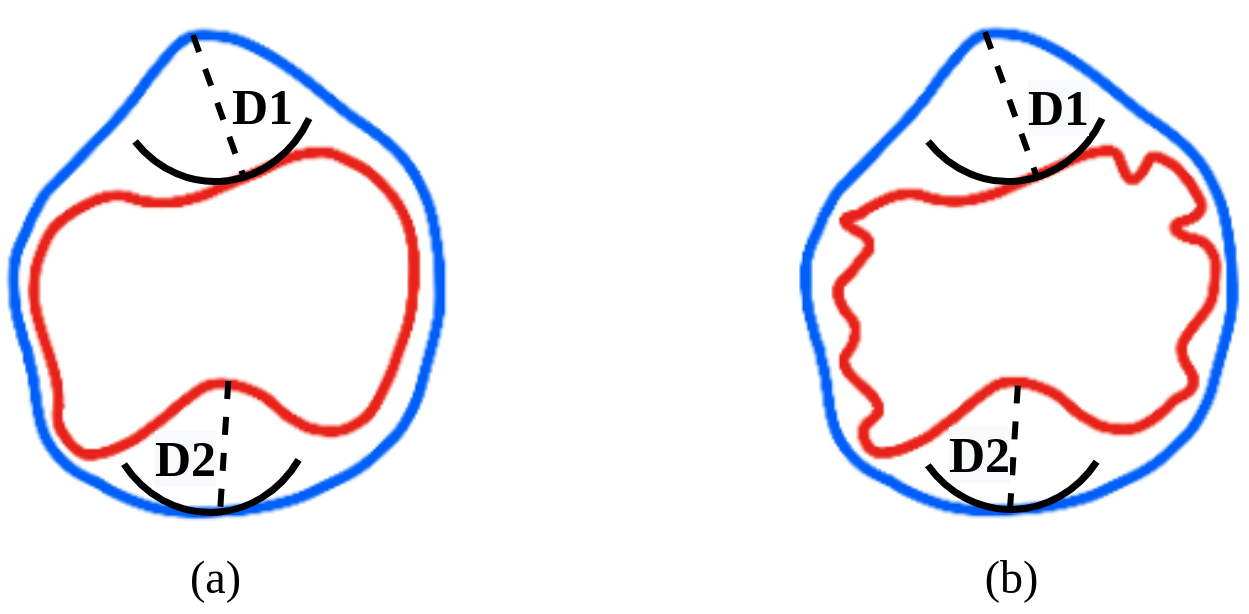}
	\caption{Illustration of HDD in two cases: smooth predicted boundary (a) and rough predicted boundary with topological changes (b). Blue curve is ground-truth boundary $G$ and red curve is predicted segmentation boundary $P$. $D_1$ is distance from $G$ to $P$ and  $D_2$ is distance from $P$ to $G$.}
	\label{Fig:HDD_Desc}
\end{figure}

Because both $P$ and $G$ are symmetric, the bidirectional Hausdorff distance between ground-truth boundary/surface $\partial G$ and the predicted segmentation boundary/surface $\partial P$ is computed as:

\begin{equation}
\operatorname{HDD} = \max{ (\max_{x \in \partial G}((|x,\partial P|_{L2})),  (\max_{x\in \partial P}((|y,\partial G|_{L2}))}
\label{eq:HDD2}
\end{equation}
Hausdorff distance, which is computed as the maximum distance between two surface, has been commonly used in practice. HDD only tells about the maximum possible distance. However, it is unable to describe the surface roughness as well as detect topological errors which are critical problems in medical imaging. Fig:\ref{Fig:HDD_Desc} illustrates some limitations of HDD. In this figure, suppose the ground-truth boundary $G$ is presented in blue curve whereas the predicted segmentation $P$ is shown in red curve. Two cases are considered in this example, namely, smooth predicted segmentation (Fig:\ref{Fig:HDD_Desc}(a)) and rough predicted segmentation (Fig:\ref{Fig:HDD_Desc}(b)) with some topological errors on the predicted segmentation boundary. Let denote $D_1$ and $D_2$ as the distance between $G$ and $P$, i.e. $D_1 = \max_{x \in \partial G}((|x,\partial P|_{L2})$ and the distance between $P$ and $G$, i.e. $D_2 = \max_{x\in \partial P}((|y,\partial G|_{L2})$. As shown in Fig:\ref{Fig:HDD_Desc}, the distance $D_1$ and $D_2$ are the same in two cases, thus the HDD is unchanged, i.e. $HDD = max (D_1, D_2)$ even the predicted boundary in Fig:\ref{Fig:HDD_Desc}(b) is different from the one in Fig:\ref{Fig:HDD_Desc}(a). Compared to the predicted boundary in Fig:\ref{Fig:HDD_Desc}(a), the one in Fig:\ref{Fig:HDD_Desc}(b) is rougher and with more topological changes.




\begin{figure}[!h]
	\includegraphics[width= 1\linewidth]{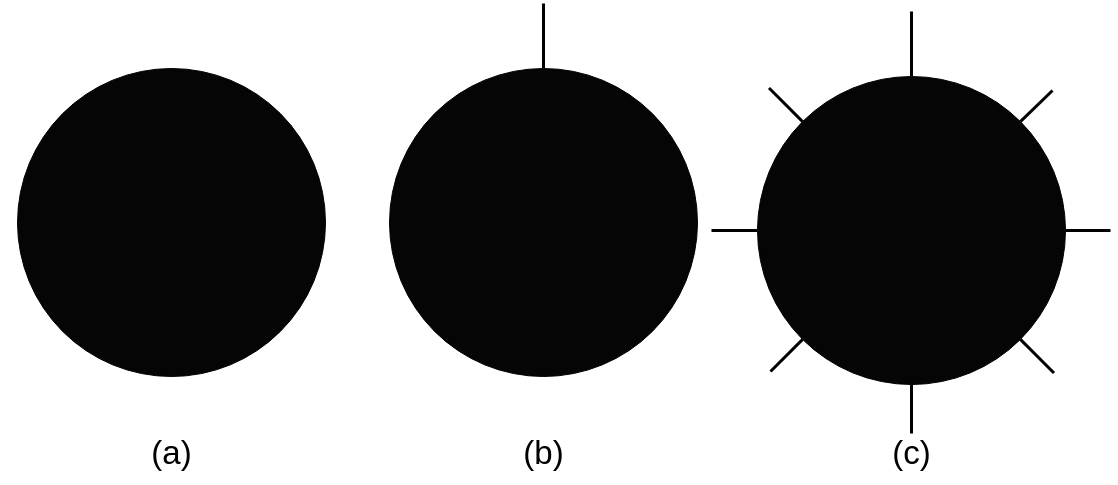}
	\caption{From left to right 2D (a): ground-truth; (b) predicted segmentation with little irregular spikes; (c) predicted segmentation with many irregular spikes.} 
	\label{Fig:HDD_Descs2}
\end{figure}

\begin{figure}[!h]
	\includegraphics[width= 1\linewidth]{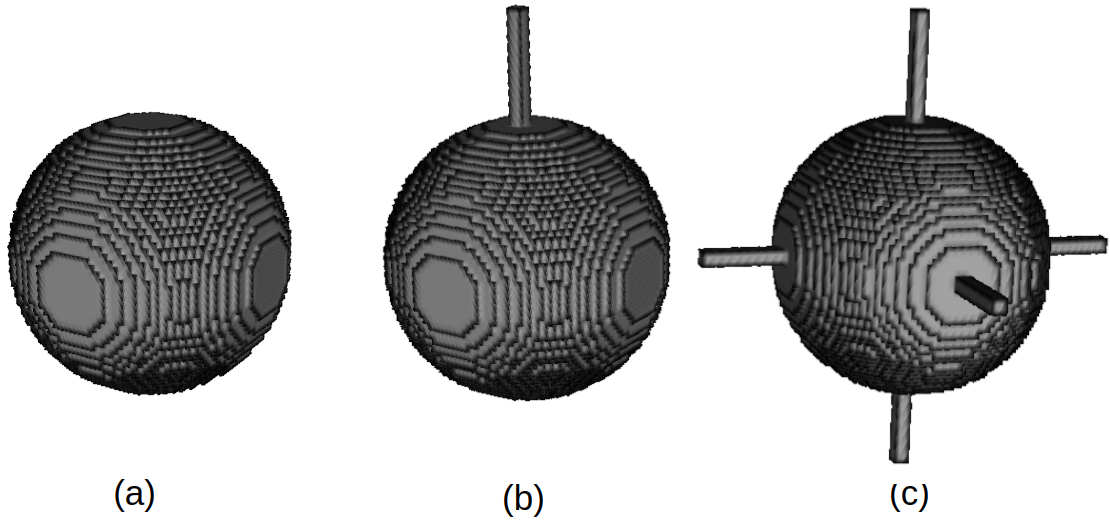}
	\caption{From left to right 3D (a): ground-truth; (b) predicted segmentation with little irregular spikes; (c) predicted segmentation with many irregular spikes.} 
	\label{Fig:3d_obj}
\end{figure}

Fig:\ref{Fig:HDD_Descs2}, \ref{Fig:3d_obj} further explains the limitations of HDD. In this example, the ground-truth is given in Fig:\ref{Fig:HDD_Descs2}, \ref{Fig:3d_obj} (a) and Fig:\ref{Fig:HDD_Descs2}, \ref{Fig:3d_obj} (b) and (c) are two different predicted segmenting results. As shown in Fig:\ref{Fig:HDD_Descs2}, \ref{Fig:3d_obj}, there are little irregular spikes on (b) and many irregular spikes on (c), however, the HDD between the ground-truth and the predicted segmenting results are the same. 

\noindent
\textbf{Average Symmetric Surface Difference (ASSD)}
ASSD \cite{chen2012domain} \cite{chen2012medical} \cite{yokota2013automated} is the average of all the distances from points/voxels on the boundary/surface of the ground-truth mask to the boundary/surface of the predicted segmentation mask, and vice versa. Denote $P$ and $G$ as the predicted segmentation mask and the ground-truth mask. The boundary/surface of $P$ and $G$ are then defined as $\partial P$ and $\partial G$. Mathematically, ASSD is computed as follows:


\begin{equation}
\operatorname{ASSD} = \frac{\sum_{x \in \partial G}(|x,\partial P|_{L2})+\sum_{x \in \partial P}(|x,\partial G|_{L2})}{|\partial G| + |\partial P|}
\end{equation}

ASSD is a good metrics for cross distance computation between boundaries of two surfaces however ASSD has the same limitations as HDD that it cannot compute the roughness or smoothness on one particular surface. 

\begin{figure}[!h]
\includegraphics[width= 1\linewidth]{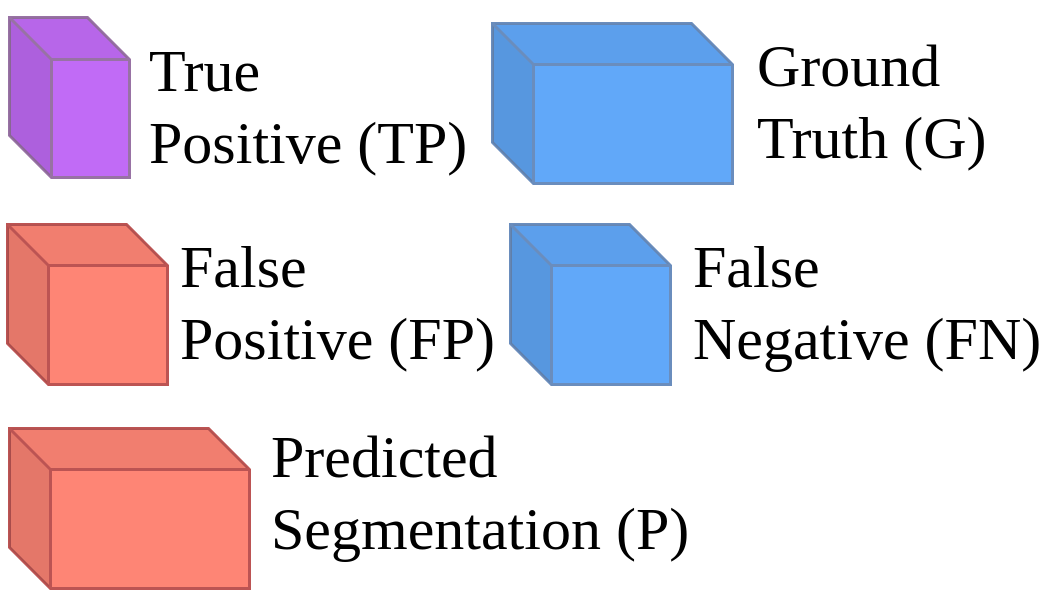}
	
\caption{Explanation of annotations \& visualization that are used in table \ref{tab:metrics}}
\label{Fig:description}
\end{figure}

The existing metrics can be summarized in Table I where the visualization is further explain in Fig.\ref{Fig:description}.

\begin{table*}[!h]
\begin{tabular}{|c|c|c|c|}
\hline
Type                          & Metrics & Equation & Visualization \\ \hline
\multirow{8}{*}{Region-Based} & DSC     &     $\frac{2|P \cap G|}{|P| + |G|}$     & \raisebox{-.5\height}{\includegraphics[height=2cm]{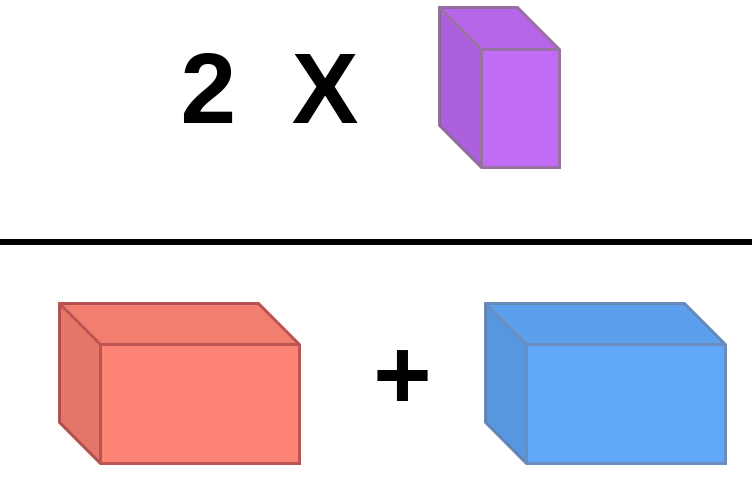}}\\ \cline{2-4} 
                              & PREC    &     $\frac{|P \cap G|}{|P|}$     & \raisebox{-.5\height}{\includegraphics[height=2cm]{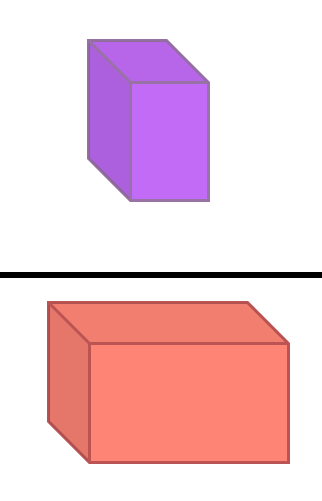}} \\ \cline{2-4} 
                              & JSC     &    $\frac{|P \cap G|}{|P \cup G|}= \frac{DSC}{2 - DSC}$      & \raisebox{-.5\height}{\includegraphics[height=2cm]{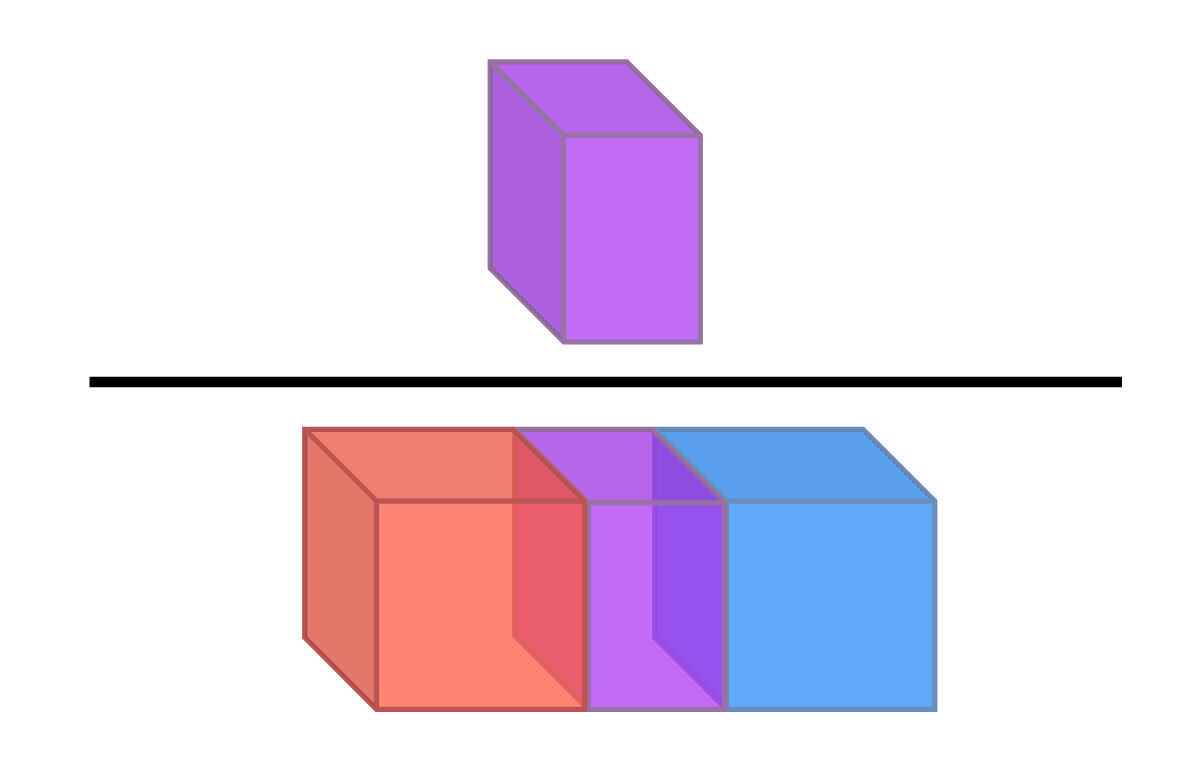}} \\ \cline{2-4} 
                              & REC, SES &    $\frac{|P \cap G|}{|G|}$      & \raisebox{-.5\height}{\includegraphics[height=2cm]{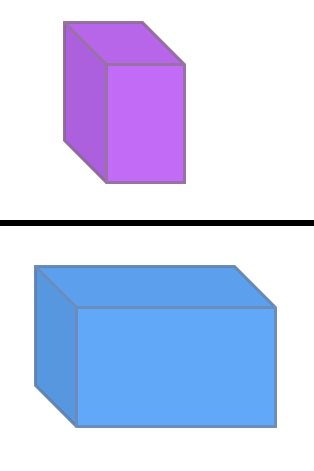}}\\ \cline{2-4} 
                              & SET      &     $\frac{|P \cup G|^C}{|G|^C}$     & \raisebox{-.5\height}{\includegraphics[height=2cm]{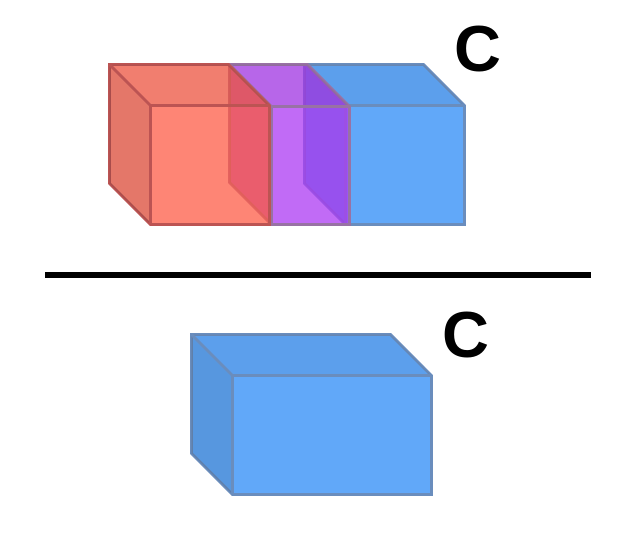}}\\ \cline{2-4} 
                              & RVD     &   $|\frac{|P| - |G|}{|G|}|$      & \raisebox{-.5\height}{\includegraphics[height=2cm]{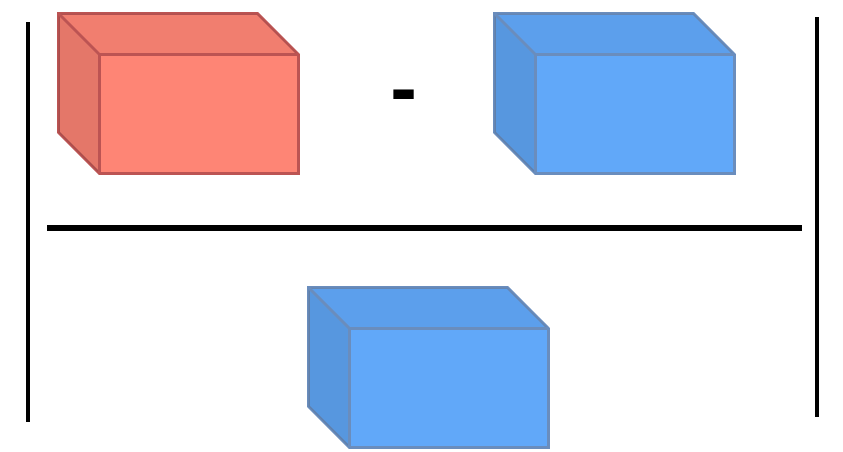}}\\ \hline
Contour Based                 & HDD     &   $\max{ (\max_{x \in \partial G}((|x,\partial P|_{L2})),  (\max_{x\in \partial P}((|y,\partial G|_{L2}))}$       & \raisebox{-.5\height}{\includegraphics[height=2cm]{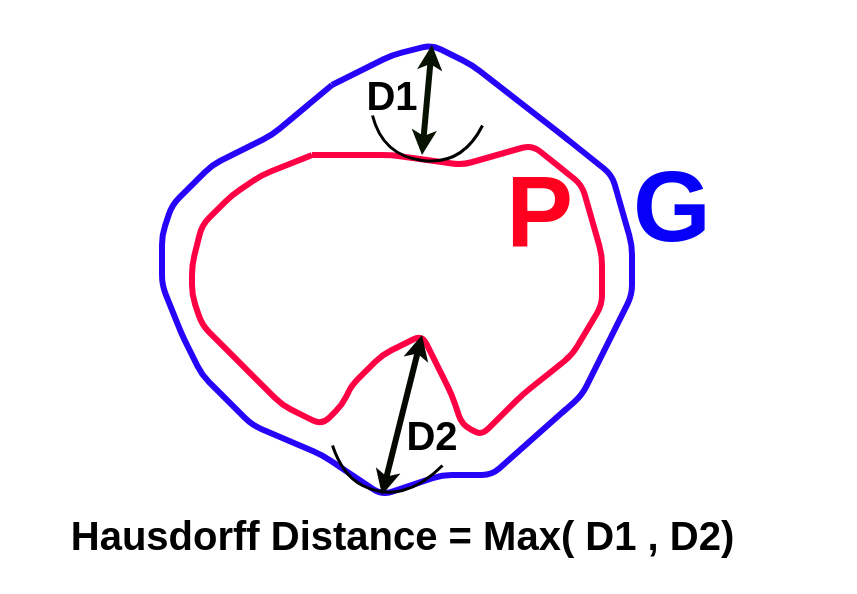}}\\ \cline{2-4} 
                            & ASSD    &   $\frac{\sum_{x \in \partial G}(|x,\partial P|_{L2})+\sum_{x \in \partial P}(|x,\partial G|_{L2})}{|\partial G| + |\partial P|}$       & \\ \hline
\end{tabular}
\caption{Summary of existing metrics on volumetric segmentation:- Red : Predicted Segmentation(P), Blue : ground-truth(G), Purple : True Positive(TP), 'C' in the subscript suggests the compliment of the vector the image description is shown in Fig \ref{Fig:description}.} 
\label{tab:metrics}
\end{table*}

\section{PROPOSED METRICS}
\label{sec:proposed}
In this section, our proposed metrics for surface roughness analysis in medical segmentation will be detailed. Our proposed roughness index and roughness distance is based on the real world average roughness parameter\cite{tonietto2019new} as described in Sec:\ref{sec:detect}. In civil engineering domain, roughness parameter of a particular surface is calculated using a laser to map the irregularities on the surface \cite{tonietto2019new}. All the symbols and notations used to describe the proposed metrics have been summarized in Table:\ref{Tab:symbols}.

\begin{table*}[!h]
\resizebox{\textwidth}{!}{%
\begin{tabular}{|l|l|l|l|} 
\hline
Symbol & Description & Symbol & Description  \\ 
\hline 
P      & Predicted segmentation mask & G       & ground-truth segmentation mask  \\ 
\hline
  $C_{0}$     & Center of gravity of a contour   &    $\partial S^w$    & Segment of an array S with a fixed window size w          \\ 
  \hline
$\zeta$ $\zeta_{i}$ $\zeta_{ij}$ $\zeta_{ijk}$        &  Distance of a contour position (i,j,k) from $C_{0}$           &  $\zeta_m$      & Distance matrix, $\zeta_m(i,j,k) = \zeta_{ijk}$   \\
\hline
 $\zeta_{Neighbor}$ & Distance $\zeta$ of neighbor & $S_{\zeta_{Neighbor}}$ & set of neighboring $\zeta_{Neighbor}$ of $\zeta$ \\ \hline $\Delta \zeta$ $\Delta \zeta_{i}$ $\Delta \zeta_{ij}$ $\Delta \zeta_{ijk}$  & Roughness at a matrix position (i,j,k)  & $\hat{\zeta}$ & Difference between $\zeta$ for P and G\\ 
 \hline 
  $\Delta \zeta_{m}$ & matrix of $\Delta \zeta$ & $\Delta \zeta_{Bm}$  & Rough Boolean matrix, where $\Delta \zeta_{Bm} \in (0,1)$   \\ 
  \hline 
\end{tabular}
}
\caption{Symbols along with their descriptions.}
\label{Tab:symbols}
\end{table*}

\subsection{Irregular Spike/Hole Detection}\label{sec:detect}

Roughness is a very important parameter that is frequently used in civil engineering domain \cite{chang2006application} \cite{tonietto2019new} \cite{gadelmawla2002roughness}. Civil engineers use the roughness parameter to measure the inconsistencies on a particular surface such as a slab of concrete or metal. A surface profile gauge or a Digital Holographic Microscope is used to map the fluctuations on the surface \cite{tonietto2019new}. The roughness parameter\cite{tonietto2019new} in civil engineering domain is defined in Eq:\ref{eq:RI_realworld} and illustrated in Fig:\ref{Fig:ria} , where $\zeta_{i}$ is the perpendicular distance of a point from the laser plane also referred to as the height coordinate \cite{tonietto2019new} is calculated using a laser moving on a fixed plane parallel to the object surface and N is the total number of points where height coordinate is calculated.

\begin{equation}
\operatorname{Roughness Parameter}  = \frac{1}{N}\sum_{i}^N{|\zeta_i|}
\label{eq:RI_realworld}
\end{equation}

We extended the term height coordinate to use it in 2D and 3D domain by calculating the distance of the surface point from the center of gravity $C_{0}$ instead of a plain, as illustrated in Fig: \ref{Fig:ric} for a closed contour laser plain can be approximated as the center of gravity. We have defined $\zeta$ (Zeta) as the distance of a surface point for a contour $P_{Surface}$ from center of gravity $C_{0}$ as shown in Eq:\ref{Eq:zetadist}. Here $P_{Surface}$ is the matrix that has value 1 or 0 based on whether the location in the segmentation mask P belongs to the surface or not respectively.

\begin{equation}
\operatorname{\zeta_{ijk}} = 
\begin{cases}
     |(i,j,k)  , C_{0}|_{L2}& P_{Surface}(i,j,k) = 1\\
    0 &  \text{Otherwise}
\end{cases}
\label{Eq:zetadist}
\end{equation}

For roughness in 2D and 3D we use a \textbf{Distance Matrix} $\zeta_m$ that contains the distance of each corresponding surface point from the center of gravity $C_{0}$ as shown in Eq:\ref{Eq:zeta}. This matrix can be used to detect and correct surface roughness. The main purpose of calculating $\zeta$ is to track the variations in surface. As illustrated in Fig:\ref{Fig:motivation} an irregular hole/spike is marked by an abrupt change in $\zeta$ while for a regular hole/spike change in $\zeta$ takes place gradually.

\begin{equation}
\operatorname{\zeta_m(i,j,k)} = \zeta_{ijk}
\label{Eq:zeta}
\end{equation}

\begin{figure}[!t]
    \includegraphics[width=1\linewidth]{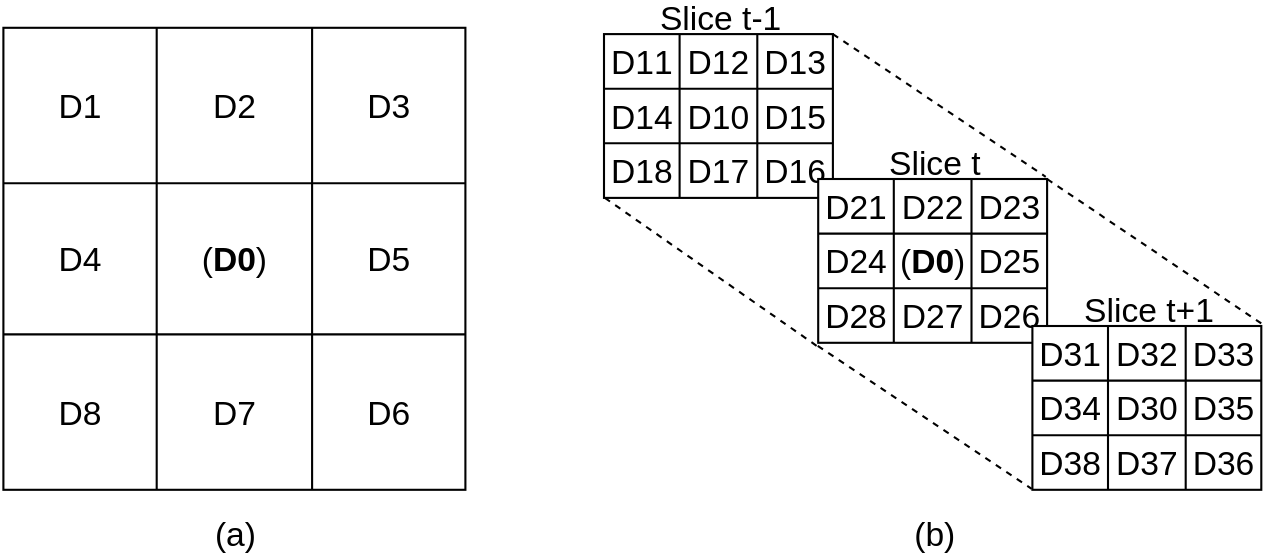}
    \caption{Illustration of how neighbors of a reference point D0 are considered for 2D (a) and 3D (b) Distance matrix $\zeta_m$ . In the given figure D0 is a position in distance matrix $\zeta_m$, that belongs to the contour and for which roughness $\Delta \zeta$ needs to be calculated.}
    \label{Fig:neb}
\end{figure}

To detect roughness we define \textbf{Roughness Matrix} $\Delta \zeta_{m}$ containing roughness value $\Delta \zeta$ (Delta zeta) for each surface location as shown in Eq:\ref{Eq:delzetam}. Roughness $\Delta \zeta$ of a location on surface  can be defined as the sum of differences between  $\zeta$ and its contour neighbors $\zeta_{Neighbors}$, belonging to the set of neighbors $S_{\zeta_{Neighbors}}$ illustrated in  Fig:\ref{Fig:neb} and described in Eq:\ref{Eq:delzeta} and Eq:\ref{Eq:rel}.  

\begin{equation}
\operatorname{\Delta \zeta_{ijk}} = \sum (\zeta_{ijk} - \zeta_{Neighbors})
\label{Eq:delzeta}
\end{equation} 

\begin{equation}
\operatorname{\zeta_{Neighbors}} \in S_{\zeta_{Neighbors}}
\label{Eq:rel}
\end{equation}

\begin{equation}
\operatorname{\Delta \zeta_m(i,j,k)} = \Delta \zeta_{ijk}
\label{Eq:delzetam}
\end{equation}

\begin{figure}[!t]
    \includegraphics[width= 1\linewidth]{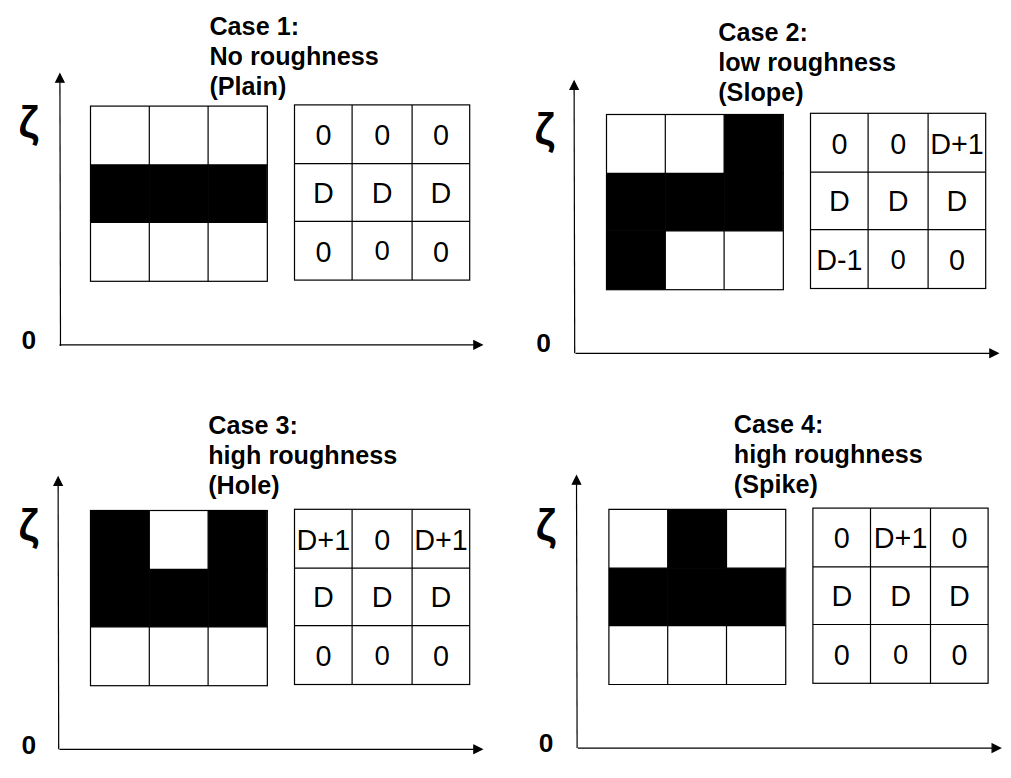}
    \caption{Different cases of roughness when dealing with 2D segmentation mask, In the figure $\zeta$ is the corresponding distance of contour point from center of gravity $C_0$}
    \label{Fig:rmi}
\end{figure}

lets consider a 2D example for various cases of roughness as shown in Fig:\ref{Fig:rmi}.

\begin{itemize}
    \item Case 1: shows the condition of a plain where the neighbors are at the same distance from $C_{0}$ as the point for which $\Delta \zeta$ needs to be calculated, so $\Delta \zeta$ will be $((D - D) + (D - D)) = 0$
    \item Case 2: shows a slope where  $\Delta \zeta$ will be $((D - D) + (D - D) + (D - (D + 1)) + (D - (D - 1))) = 0$.
    \item Case 3: is an example of hole where  $\Delta \zeta$ will be $((D - D) + (D - D) + (D - (D + 1)) + (D - (D + 1))) = 2$
    \item Case 4: which is a spike where $\Delta \zeta$ will be $((D - D) + (D - D) + (D - (D + 1))) = 1$
\end{itemize}

Hence it can be easily concluded that $|\Delta \zeta|$ for a location close to zero will denote a smooth surface and greater then zero will refer to rough surface.

\subsection{Roughness Metrics}\label{sec:ri}

\begin{figure}
\centering
\begin{subfigure}{1\textwidth}
  \centering
  \includegraphics[width=1.0\linewidth]{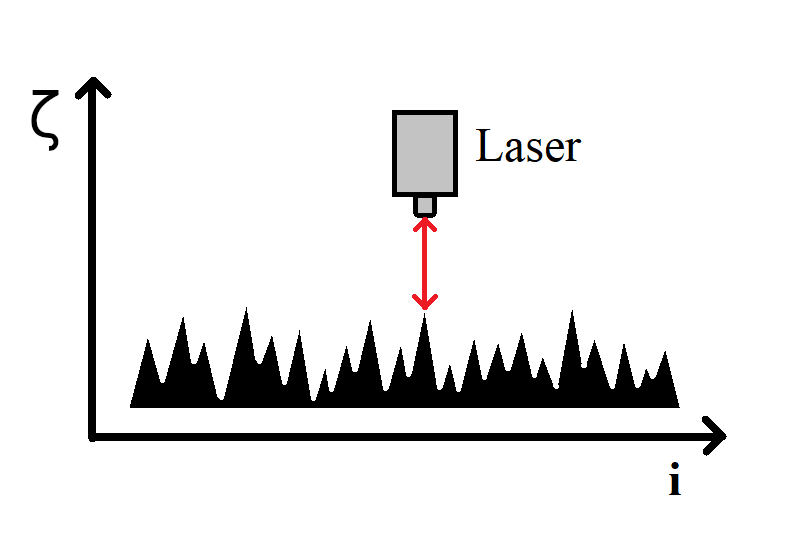}
  \caption{In civil engineering domain Roughness is calculated by moving a laser parallel to the surface to find height coordinate $\zeta$ and then using it to find the roughness parameter of the surface using Eq:\ref{eq:RI_realworld}.}
  \label{Fig:ria}
\end{subfigure}
\begin{subfigure}{1\textwidth}
  \centering
  \includegraphics[width=1.0\linewidth]{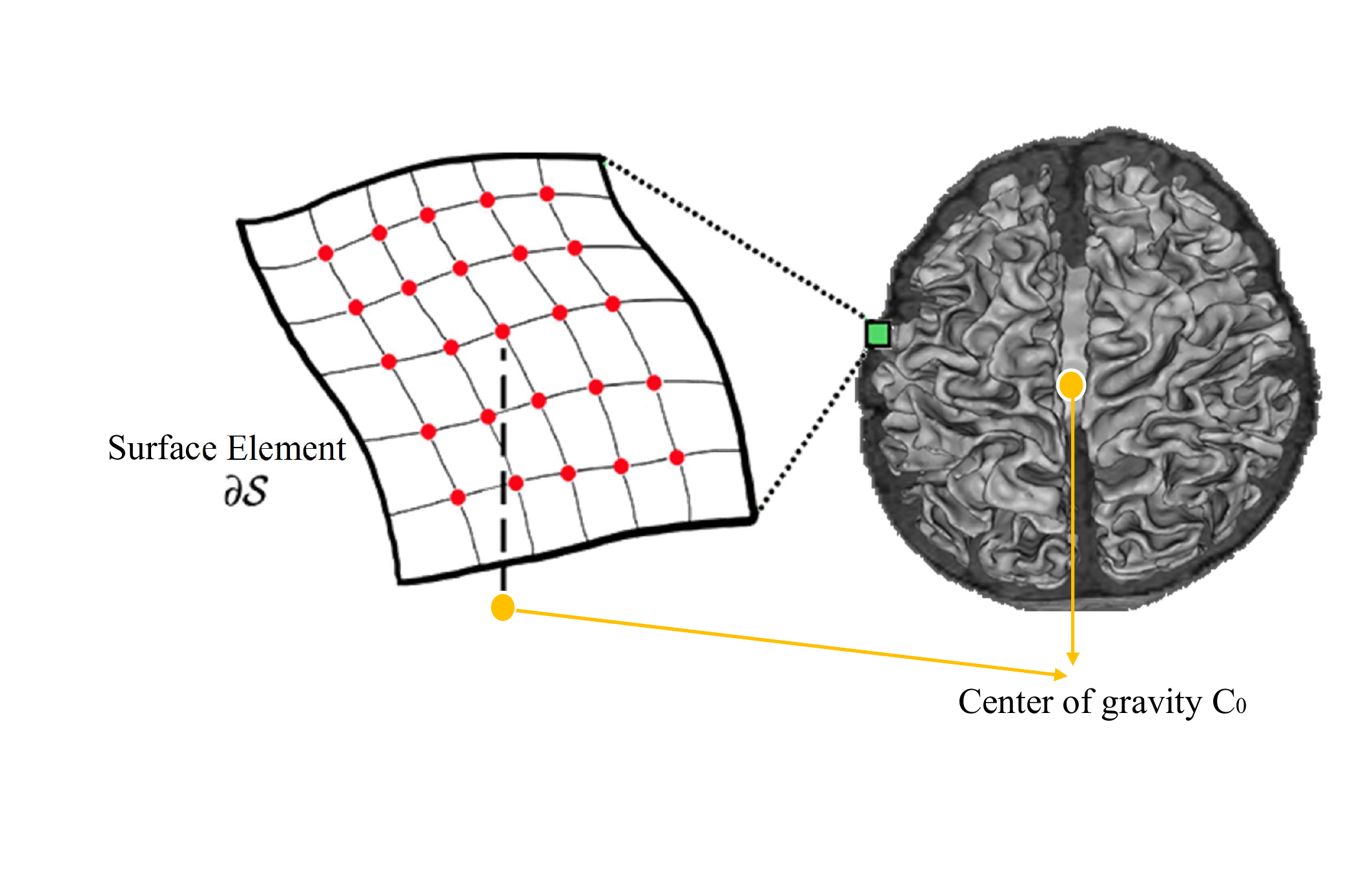}
  \caption{In medical domain the segmentation mask is a closed object as opposed to a flat surface in civil engineering, hence we calculate the distance of object surface $\zeta$ from center of gravity instead of a plane. This $\zeta$ is used to calculate roughness index and surface/roughness distance.}
  \label{Fig:rib}
\end{subfigure}
\begin{subfigure}{1.0\textwidth}
\includegraphics[width= 1.0\linewidth]{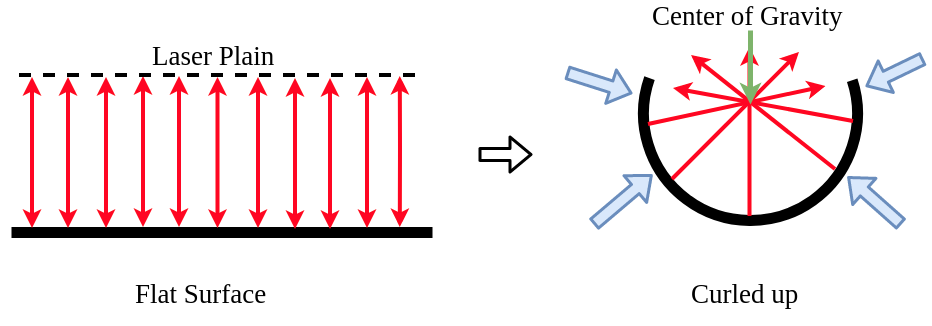}
\caption{Laser plane can be approximated to center of gravity if the surface is rolled into a closed contour.}  
\label{Fig:ric}
\end{subfigure}
\caption{Illustration of how to compute roughness (a): Roughness parameter is calculated by moving laser on fixed plane paralleled to the surface, (b): Roughness index and surface/roughness distance is calculate from an origin which is defined as the center of medical image and (c): How the method of calculating roughness can be extended to medical imaging domain.}
\end{figure}

Roughness parameter is a term usually used to determine the roughness of a solid surfaces\cite{chang2006application} \cite{tonietto2019new} \cite{gadelmawla2002roughness}. We extended this term to use in 2D and 3D surface vector domain. As shown in Eq:\ref{eq:RIndex} The \textbf{Roughness Index} (RI) in 3D can be calculated by dividing the segmentation surface S into small surface element $\partial S^w$ of a fixed window size w, and then calculating the average deviation of $\zeta$ from the mean $\zeta_{Mean}$ for all surface voxels in the surface element $\partial S^w$ as illustrated in Fig:\ref{Fig:rib}. In Eq:\ref{eq:RIndex} $\partial S_i^w$ denotes a point on the surface element $\partial S^w$, M is the total no of surface elements $\partial S^w$ that the contour surface is divided into and N is the total no of points $i$ inside each surface element that belong to the contour surface, Here $|\partial S_i^w, C_0|_{L2}$ is equal to $\zeta$ that we calculated in previous section.

\begin{equation}
\small
\operatorname{RI}  = \frac{1}{M}\sum_{ \partial S^{w} \in S}^{M}{\frac{1}{N}\sum_{i}^N{|(|\partial S_{i}^{w},C_{0}|_{L2})- Mean(|\partial S_{i}^{w}, C_{0}|_{L2})|}}
%
\label{eq:RIndex}
\end{equation}


Roughness is a relative quantity. An object that is rough as compared to one surface may be smooth as compared to other.  Hence it can be difficult to tell about the roughness of a surface unless we have a baseline to compare the roughness index. Hence we introduce \textbf{Roughness Ratio} (RR) that tells about the relative difference between roughness of two objects. The roughness ratio has been defined in Eq: \ref{eq:RR} where $\mathcal{RI}_{P}$ and $\mathcal{RI}_{G}$ are the roughness index of predicted segmentation and ground-truth respectively.

\begin{equation}
\operatorname{RR}  = \frac{|RI_P - RI_G|}{RI_G}
\label{eq:RR}
\end{equation}

\subsection{Roughness Distance}\label{sec:surfdist}


In this section, we propose \textbf{Roughness Distance} which is considered as surface distance between two surfaces. Let denote $\hat{\zeta}$ as the difference between the $\zeta$ for predicted segmentation ($\zeta_P$) and ground-truth segmentation ($\zeta_G$) as shown in Eq:\ref{Eq:crossdelzeta} and \textbf{Roughness Distance Matrix} $\hat{\zeta}_m$ as the matrix containing $\hat{\zeta}$ values as shown in equation Eq:\ref{Eq:crossdelzetam}


\begin{equation}
\operatorname{\hat{\zeta}_{ijk}} = \zeta_{P_{ijk}} - \zeta_{G_{ijk}}
\label{Eq:crossdelzeta}
\end{equation}

\begin{equation}
\operatorname{\hat{\zeta}_{m}(i,j,k)} = \hat{\zeta}_{ijk}
\label{Eq:crossdelzetam}
\end{equation}

Simply speaking roughness distance matrix $\hat{\zeta}_m$ can be calculated by subtracting distance matrix for ground-truth segmentation $\zeta_{mG}$ from distance matrix for predicted segmentation $\zeta_{mP}$ as shown in equation Eq:\ref{Eq:crossdelzetamm}. Roughness distance can be used to calculate the roughness change between two object, ground-truth and predicted segmentation in our case.

\begin{equation}
\operatorname{\hat{\zeta}_{m}} = \zeta_{mP} - \zeta_{mG}
\label{Eq:crossdelzetamm}
\end{equation}

We also propose \textbf{Average Roughness Distance} (ARD) which as the name suggest is the average surface/roughness distance between two objects as shown in Eq:\ref{Eq:ard}. ARD is a metric that tells us about the average difference between the surface of two objects.  ARD can be used as a substitute of HDD to compare roughness. 

\begin{equation}
\operatorname{ARD} = Mean(|\hat{\zeta}_{m}|)
\label{Eq:ard}
\end{equation}

\subsection{Surface Smoothing}\label{sec:smooth}


In this section we will propose a method for smoothing a contour that has roughness on its surface. Smooth contour can be obtained by using either of the two methods which include using roughness matrix $\Delta \zeta_{m}$ or roughness distance matrix $\hat{\zeta}_m$ that were calculated in previous sections   

For contour smoothing rough boolean matrix $\Delta \zeta_{Bm}$ is used where the value can be one or zero based on whether the position is considered as rough or smooth respectively as shown in Eq:\ref{Eq:delzetabm} where $\kappa$ is the threshold roughness value in range of (0 , $\max{(\Delta \zeta_m)}$).

\begin{equation}
 \operatorname{\Delta \zeta_{Bm(i,j,k)}} =  
\begin{cases}
    1,& |\Delta \zeta_{ijk}| > \kappa\\
    0, &  \text{Otherwise}
\end{cases}
\label{Eq:delzetabm}
\end{equation}

Similarly $\Delta \zeta_{Bm}$ can also be computed using the roughness distance matrix $\hat{\zeta}_m$ as shown in Eq:\ref{Eq:delzetabmcross} where $\kappa_{c}$ is the threshold distance in range of (0 , $\max{(\hat{\zeta}_m)}$).

\begin{equation}
 \operatorname{\Delta \zeta_{Bm(i,j,k)}} =  
\begin{cases}
    1,& |\hat{\zeta}_{ijk}| > \kappa_{c}\\
    0, &  \text{Otherwise}
\end{cases}
\label{Eq:delzetabmcross}
\end{equation}

This rough boolean matrix $\Delta \zeta_{Bm}$ can be used for contour smoothing as shown in Eq: \ref{Eq:rect} where $P_{Rough}$ is the segmentation mask before contour smoothing and $P_{Smooth}$ is the one after smoothing. It is important to note here that both methods return a smooth contour, However while $\Delta \zeta_m$ requires only the rough segmentation contour, $\hat{\zeta}_m$ also required the corresponding ground-truth segmentation contour.

\begin{equation}
\operatorname{P_{Smooth}} = |P_{Rough} - \Delta \zeta_{Bm}|
\label{Eq:rect}
\end{equation}

\begin{algorithm}
\caption{Calculate Roughness index(RI) of a 3D contour}
\begin{algorithmic}
\REQUIRE $ S \in \{0,1\} \vee S[x,y,z]  $
\STATE $(X_0,Y_0,Z_0) \leftarrow 0$
\STATE $N \leftarrow 0 $
\STATE $RI \leftarrow 0 $
\FOR{$ (X,Y,Z)  \in  S $}

\STATE $(X_0,Y_0,Z_0) \leftarrow (X_0,Y_0,Z_0) + (X,Y,Z)$
\STATE $N \leftarrow N + 1 $
\ENDFOR
\STATE $(X_0,Y_0,Z_0) \leftarrow (X_0,Y_0,Z_0) / N$
\STATE $M \leftarrow 0 $
\FOR{$ \partial S^w \in S $}

\STATE $C \leftarrow 0 $
\FOR{$ (X,Y,Z)  \in  \partial S^w $}

\STATE $D_0 \leftarrow D_0 + | (X,Y,Z) , (X_0,Y_0,Z_0)|_{L2} $
\STATE $C \leftarrow C + 1 $
\ENDFOR
\STATE $D_0 \leftarrow D_0 / C$
\STATE $N \leftarrow 0 $
\STATE $R_{s} \leftarrow 0 $
\FOR{$ (X,Y,Z)  \in  \partial S^w $}
\STATE $R_{s} \leftarrow R_{s} + |(|(X,Y,Z) , (X_0,Y_0,Z_0)|_{L2}) - D_0| $
\STATE $N \leftarrow N + 1 $
\ENDFOR
\STATE $R_{s} \leftarrow R_{s}/N $
\STATE $M \leftarrow M + 1 $
\STATE $RI \leftarrow RI + R_{s} $
\ENDFOR
\STATE $RI \leftarrow RI/M $
\end{algorithmic}
\end{algorithm}

\section{DISCUSSION \& EXPERIMENTATION}
\label{sec:exp}
\subsection{Results and comparison}

The Table \ref{tab:metrics} summarizes all the metrics that are currently being used for evaluation of 3D medical images. As stated earlier the region based metrics cannot calculate the roughness and smoothness of a 3D contour. Also Hausdorff distance is capable of finding the maximum distance between the ground-truth and predicted but it fails to capture the small roughness on the surface.

In our experiments we used 2D and 3D segmentation images of size (100 $\times$ 100) and (100 $\times$ 100 $\times$ 100) as shown in Fig:\ref{Fig:HDD_Descs2} and Fig:\ref{Fig:3d_obj} respectively. In both cases (a) is a smooth segmentation treated as ground-truth, (b) is a segmentation with small spike of 20 pixel length and (c) has many spikes, where the top spike has the same length of 20 as (b) and all other spikes are of smaller length for both 2D and 3D example.

Mathematically the roughness index ($RI$) for a circle and sphere should be 0, but because for images the coordinate system is integral not continuous, even a smooth circle has a small $RI$ greater then 0. Hence, we treat it as Residual Roughness Index $RI_{Residual}$. In our experiments we have treated the RI for ground-truth segmentation as residual. The absolute roughness index $RI_{Absolute}$ can be calculated by subtracting $RI_{Residual}$ from $RI$ as shown in Eq:\ref{eq:riabs} .

\begin{equation}
\operatorname{RI_{Absolute}}  = RI - RI_{Residual}
\label{eq:riabs}
\end{equation}

\begin{figure}[!t]
\includegraphics[width= 1\linewidth]{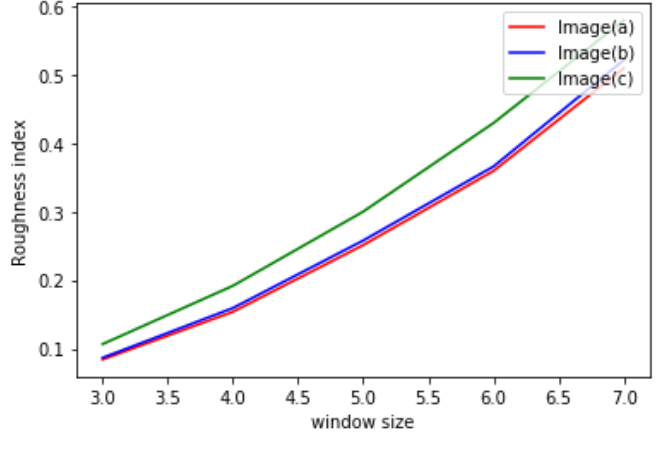}
\caption{Graph of Roughness index Vs Window size for 2D images in Fig:\ref{Fig:HDD_Descs2}(a), (b) and (c).}
\label{Fig:graph2d}
\end{figure}

\begin{figure}[!t]
\includegraphics[width= 1\linewidth]{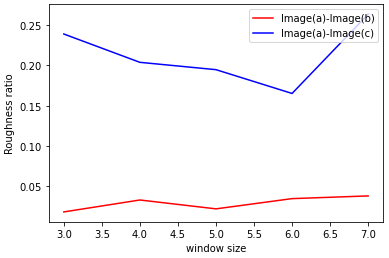}

\caption{Plot of Roughness ratio Vs window size for Fig:\ref{Fig:HDD_Descs2}(a)(ground-truth), Fig:\ref{Fig:HDD_Descs2}(b)(little roughness predicted segmentation) and Fig:\ref{Fig:HDD_Descs2}(a)(ground-truth segmentation), Fig:\ref{Fig:HDD_Descs2}(c)(high roughness predicted segmentation)}
\label{Fig:rr2d2}

\end{figure}

\begin{figure}[!t]
\includegraphics[width= 1\linewidth]{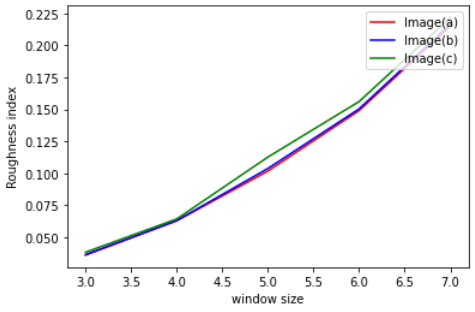}
\caption{Graph of Roughness index Vs Window size for 3D images in Fig:\ref{Fig:3d_obj}(a), (b) and (c).}
\label{Fig:graph3d}
\end{figure}

\begin{figure}[!t]
\includegraphics[width= 1\linewidth]{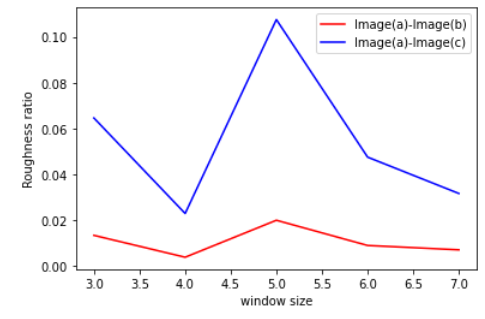}

\caption{Plot of Roughness ratio Vs window size for Fig:\ref{Fig:3d_obj}(a)(ground-truth), Fig:\ref{Fig:3d_obj}(b)(little roughness predicted segmentation) and Fig:\ref{Fig:3d_obj}(a)(ground-truth segmentation), Fig:\ref{Fig:3d_obj}(c)(high roughness predicted segmentation)}
\label{Fig:rr3d2}
\end{figure}

In our experiments we varied the window size and calculated roughness index for various window sizes. For 2D image we first performed 2D convolution on the image using a kernel and dilation operator to convert the 2D image into a contour. We then used a 2D window and moved it on the image and calculated the variation of distance from each boundary location to the center of gravity of the contour using mean distance for all boundary locations present in the window. We used the s

trides equal to the window size so that each location is used to calculate roughness index exactly once. The plot of $RI$ and $RR$ vs window size for images in Fig:\ref{Fig:HDD_Descs2} is shown in Fig:\ref{Fig:graph2d} and Fig:\ref{Fig:rr2d2} respectively. Similarly for 3D images we first performed 3D convolution on the image using a kernel and dilation operator to convert the 3D image into a contour. We then used a 3D window to calculate the $RI$ of each image in Fig:\ref{Fig:3d_obj}. The $RI$ and $RR$ vs window size graph has been shown in Fig:\ref{Fig:graph3d} and Fig:\ref{Fig:rr3d2} respectively.

It can be inferred from Fig:\ref{Fig:graph2d} and Fig:\ref{Fig:rr2d2} that window size plays a very important role in RI calculation. In our experimentation we found that the optimal window size must be between 3\% to 10\% of the image smallest dimension. It is also important to note that Roughness index is a standalone metrics but it can be used to compare the roughness of two image through roughness ratio that we have used in our experiments. 


A tabular comparison between roughness ratio ($RR$), average roughness distance ($ARD$) and Hausdorff distance ($HDD$) has been shown in Table \ref{tab:comp}. It can be easily inferred from the table that $RR$ and $ARD$ were capable of finding the difference in roughness for the two image pairs that $HDD$ failed to do.

\begin{table*}[!h]
\resizebox{\textwidth}{!}{%
\begin{tabular}{|l|l|l|l|l|}
\hline
Images & \begin{tabular}[c]{@{}l@{}}Absolute\\ Roughness \\ Index\end{tabular}   & \begin{tabular}[c]{@{}l@{}}Roughness \\ Ratio\end{tabular} &\begin{tabular}[c]{@{}l@{}}Average\\ Roughness \\Distance\end{tabular}& \begin{tabular}[c]{@{}l@{}}Hausdorff \\ Distance\end{tabular} \\ \hline
 \includegraphics[width= 0.21\textwidth]{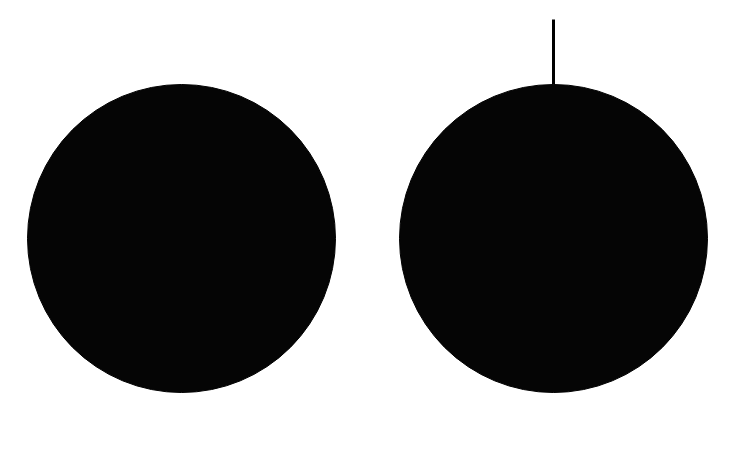}      & 0 and 0.0120 & 0.0235 &  0.3178       & 20                 \\ \hline
 \includegraphics[width= 0.21\textwidth]{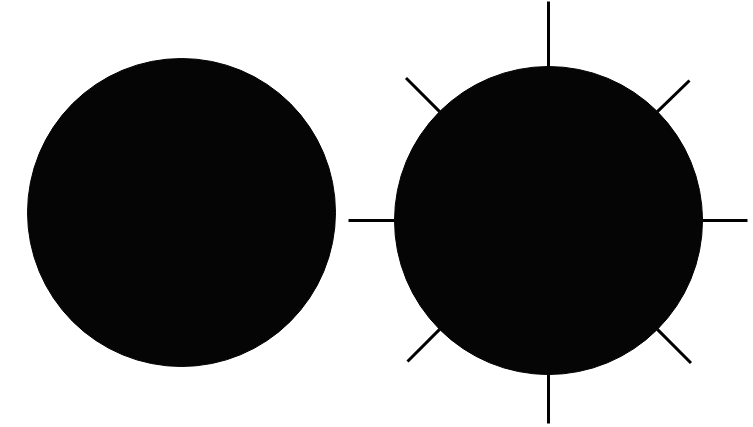}      & 0 and 0.0703 & 0.1377 &  0.7736       & 20                 \\ \hline
 
  \includegraphics[width= 0.21\textwidth]{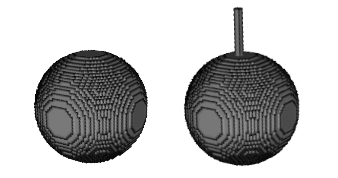}      & 0 and 0.0015 & 0.0070 &  0.0592       & 20                 \\ \hline
 \includegraphics[width= 0.21\textwidth]{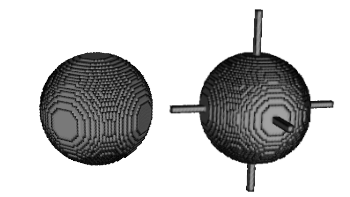}      & 0 and 0.0068 & 0.0317 &  0.0692       & 20                 \\ \hline
\end{tabular}%
}

\caption{Comparison of Roughness Index RI, Roughness Ratio RR and Average Roughness Distance ARD with respect to Hausdorff Distance, The window size considered for RI calculation is 7\% of image size i.e. 7. From top to bottom Fig:\ref{Fig:HDD_Descs2}(a)(2D ground-truth segmentation) and  Fig:\ref{Fig:HDD_Descs2}(b)(2D little roughness predicted segmentation), Fig:\ref{Fig:HDD_Descs2}(a)(2D ground-truth segmentation) and Fig:\ref{Fig:HDD_Descs2}(c)(2D high roughness predicted segmentation), Fig:\ref{Fig:3d_obj}(a)(3D ground-truth segmentation) and  Fig:\ref{Fig:3d_obj}(b)(3D little roughness predicted segmentation), Fig:\ref{Fig:3d_obj}(a)(3D ground-truth segmentation) and Fig:\ref{Fig:3d_obj}(c)(3D high roughness predicted segmentation)} 

\label{tab:comp}
\end{table*}

\begin{figure}[!t]
\includegraphics[width= 1\linewidth]{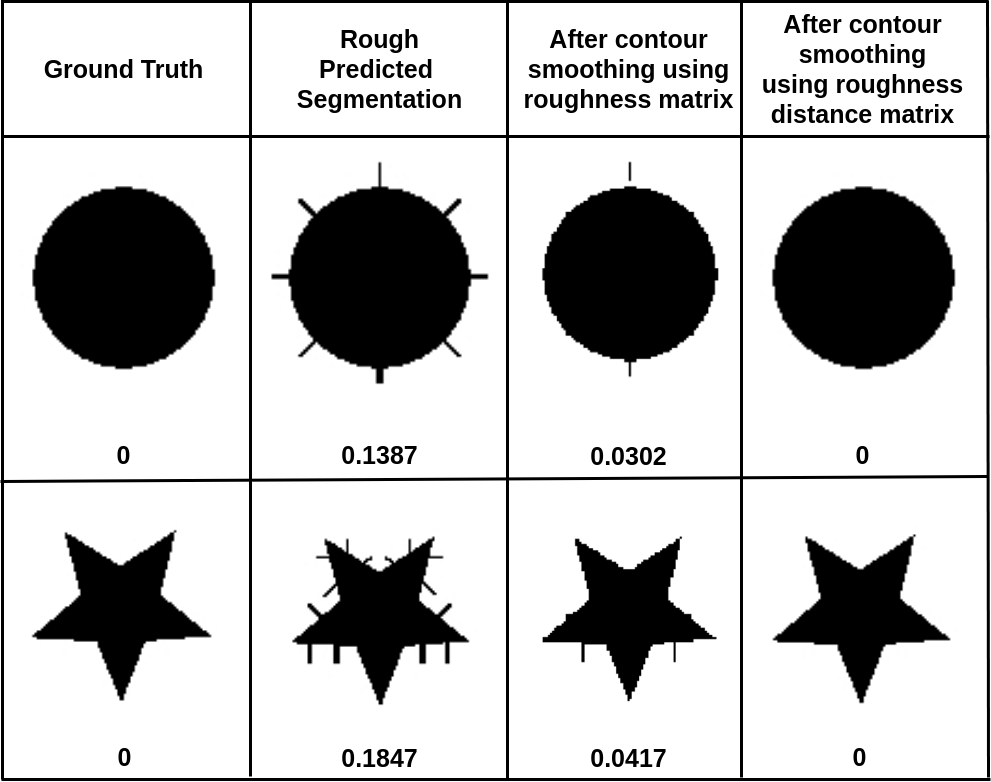}

\caption{Contour smoothing for two type of segmentation, Circle (top) and star (bottom). Smoothing has been performed using the methods discussed in sections Sec:\ref{sec:detect} and Sec:\ref{sec:surfdist} and the Absolute roughness index $RI_{Absolute}$ is specified below each image. The window size for RI calculation is 7\% of image size i.e. 7 and roughness threshold $\kappa$ , $\kappa_{c}$ for the given experiments was taken as 0.}
\label{Fig:exps}
\end{figure}

For contour smoothing we used the algorithms discussed in Sec:\ref{sec:detect} and Sec:\ref{sec:surfdist} to smooth the contours as shown in Fig:\ref{Fig:exps}. It is clear from Fig:\ref{Fig:exps} that roughness distance method produces a more satisfying result as compared to roughness matrix method, The reason being that roughness distance uses ground-truth segmentation as reference. However this is also a drawback because roughness distance method is constraint by the need of a reference distance matrix. Furthermore roughness distance method will produce unsatisfactory results if the center of gravity for P and G are not same, i.e. the segmentation masks are not aligned. However this problem can be overcome by using the center of gravity of segmentation mask G for both P and G.

Roughness matrix method is a robust method for detecting and removing surface roughness. For roughness calculation we considered a window size of three which includes a total of eight neighbors for 2D and twenty six neighbors for 3D as shown in Fig:\ref{Fig:neb}. this method is capable for detecting irregular spikes of width 1 pixel, However this method can be extended to detect spikes of multiple pixels by increasing the window size and the number of neighbors in neighbors set $S_{\zeta_{Neighbors}}$. The method for smoothing holes is same but in that case we will add a surface point to the surface instead of removing it in case of a spike.

\section{CONCLUSION}
In this paper we first discussed the pros and cons of various metrics that have been commonly used for the medical image segmentation task. We emphasize more on the limitations of existing metrics for volumetric segmentation. We then proposed (i) an algorithm that helps to detect all irregular spikes/holes that exist in the object surface; (ii) a roughness metric that describes how rough of a given object; (iii) a roughness distance that aims at comparing the surfaces between two given objects; (iv) an algorithm that aims at removing irregular spikes/holes to smooth the surface. Compare to other volumetric segmentation metrics i.e. Hausdorff distance, our proposed roughness distance is able to measure the topological error whereas roughness metric present the surface roughness. Furthermore, our proposed irregular spikes/holes detection and surface smoothing can be applied as a post-processing step in any image segmentation algorithm to improve the accuracy.




\section*{ACKNOWLEDGEMENT}
This research was supported in part by the Department of Radiology, University of Arkansas of Medical Science UAMS,

\bibliography{Example} 
\bibliographystyle{apalike} 

\end{document}